# Effects of the applied fields' strength on the plasma behavior and processes in E×B plasma discharges of various propellants: I. Electric field

M. Reza*[1], F. Faraji*, A. Knoll*

*Plasma Propulsion Laboratory, Department of Aeronautics, Imperial College London, London, United Kingdom

**Abstract**: We present, in this two-part article, an extensive study on the influence that the magnitudes of the applied electric ($E$) and magnetic ($B$) fields have on a collisionless plasma discharge of xenon, krypton, and argon in a 2D radial-azimuthal configuration with perpendicular orientation of the fields. The dependency of the behavior and the underlying processes of E×B discharges on the strength of electromagnetic field and ion mass has not yet been studied in depth and in a manner that can distinguish the role of each individual factor. This has been, on the one hand, due to the significant computational cost of conventional high-fidelity particle-in-cell (PIC) codes that do not allow for extensive simulations over a broad parameter space within practical timeframes. On the other hand, the experimental efforts have been limited, in part, by the measurements' spatial and temporal resolution as well as the ability to independently control individual discharge parameters. In this sense, the notably reduced computational cost of the reduced-order PIC scheme enables to numerically cast light on the parametric variations in various aspects of the physics of E×B discharges, such as high resolution spatial-temporal mappings of the plasma instabilities, and to inform the design of experimental campaigns. In part I of the article, we focus on the effects of the $E$-field intensity. We demonstrate that the intensity of the field determines two distinct plasma regimes, which are characterized by different dominant instability campaigns. At relatively low $E$-field magnitudes, the Modified Two Stream Instability (MTSI) is dominant, whereas, at relatively high $E$-field magnitudes, the MTSI is mitigated, and the Electron Cyclotron Drift Instability (ECDI) becomes dominant. These two regimes are identified for all studied propellants. Consequent to the change in the plasma regime, the radial distribution of the axial electron current density and the electron temperature anisotropy vary.

## Section 1: Introduction

Partially magnetized plasmas in a perpendicular configuration of the electric and magnetic fields have found important applications across the industries over the last decades, from magnetrons for plasma-assisted manufacturing to Hall thrusters for in-space plasma propulsion. These "E×B" plasma discharges have been also the subject of extensive scientific research due to the intriguing and rich nature of the underlying physics that, in part, resemble the phenomena occurring in other areas of applied plasma physics, such as the fusion energy. Fully understanding the characteristics and the interactions of the plasma phenomena in E×B discharges can open the door to the design of more efficient plasma technologies, enable the establishment of first-principles predictive numerical models that can facilitate the development of the technologies by reducing the reliance on empirical, test-dependent practices, and unlock novel technological solutions.

Nonetheless, as evidenced from the large body of previous research in the literature [1]-[5], the multifaceted and highly coupled nature of the underlying plasma processes, such as instabilities and particles' transport, in the E×B discharges points to the fact that, to gain a comprehensive knowledge of the plasma phenomena, in part by untangling the role of each individual physical factor and/or mechanism, extensive high-fidelity parametric investigations are necessary over a broad parameter space of plasma conditions. Ideally, such studies shall be carried out using three-dimensional kinetic plasma simulations that can capture the 3D interactions and dependencies of the phenomena [4][5] and are of sufficiently high fidelity. The important role of the experiments shall not be overlooked as well. The experimental characterizations can enable the validation of the numerical observations and provide complementary insights from real-world devices and settings. However, the experiments also have limitations in terms of the physical insights they can provide. These primarily root in factors such as the probes' accessibility within often constrained geometries of E×B devices, the spatial and temporal resolution of the measurements as well as noise, and the sensitivity of the sensors to plasma conditions and environmental interactions. This emphasizes the necessity for the computational and experimental endeavors to go hand-in-hand toward revealing the remaining mysteries behind the physics of E×B discharges.

Focusing on the computational component in light of the context of our article, despite the promising efforts carried out so far using various approaches to enable cost-efficient 3D kinetic simulations that can be applicable for practical parametric studies of plasma systems [6]-[10], this numerical capability still do not exist. As a result, currently, one may resort to 2D simulations that, although do not provide the full-3D picture of the involved phenomena, they can still help advance our knowledge of the dependencies and variations in the plasma behavior due to different factors.

---
[1] **Corresponding Author** (m.reza20@imperial.ac.uk)

In any case, even traditional 2D kinetic simulations, such as those based on the particle-in-cell (PIC) method [11], exhibit enormous computational resource demand [12] which makes extensive parametric studies using these tools unfeasible over practical timeframes. The reduced-order PIC scheme [6][13]-[15], that features several factors reduction in the computational cost for high-fidelity simulations that reproduce the results of the conventional multi-dimensional PIC codes [6][16], serves as a viable enabler to allow extended high-fidelity studies of plasma configurations over a wide set of plasma parameters. An example of such parametric analyses were presented in our previous publication [17].

Adopting a Hall-thruster-representative plasma discharge as the proving ground, we discuss, in this two-part article, the effects that the magnitudes of the electromagnetic field have on the plasma behavior and its underlying processes. We use a 2D configuration representing the radial-azimuthal cross-section of a Hall thruster. Hence, the results of the parametric assessments in this article build up particularly on the insights derived from the investigations reported in Refs. [17][18]. An additionally important aspect of this article is that it evaluates the effects of the electromagnetic fields' strength for various propellants (gases) of industrial relevance for E×B plasma technologies, namely, xenon, krypton, and argon.

In this part I of the article, we focus on the effects of the intensity of the axial electric field. It is worth pointing out that the influences of the $E$-field magnitude on the phenomena in the E×B discharges, such as the electrons' transport, plasma species composition, and the plasma sheath characteristics, have been experimentally assessed in a number of prior works [19]-[22]. However, these investigations did not focus on the characteristics of the underlying small-scale and/or kinetic instabilities and were mostly carried out for a single gas (propellant).

We numerically assess in this paper the impacts of the $E$-field's intensity across the plasma discharges of the three mentioned propellants in two respects: (i) the development of the Electron Cyclotron Drift Instability (ECDI) and the Modified Two Stream Instability (MTSI), (ii) the characteristics of these instability modes and their influence on the plasma behavior. It is noteworthy that the axial electric field intensity can be thought of as an operational parameter of a Hall thruster, which changes spatially within the thruster's domain with especially strong axial gradients [1][18][23], but also dynamically in time within the thruster's breathing cycles [1][12][18]. Thus, our studies here may amount to analyzing the influences that these variations in the electric field can have on the plasma.

The ECDI has been studied extensively using theoretical, numerical, and experimental means over the past few decades, and there are numerous relevant articles on this instability mode, for instance Refs. [1][5][24]-[30]. The MTSI has also gathered the interest of the scientific E×B plasma community in the recent years, resulting in several publications that aimed to investigate its characteristics and interactions with other instability modes such as the ECDI [5][17][31]-[33]. Based on the available literature, the main outstanding questions with regard to these two instability modes currently surround the following: (a) the conditions for their excitation in various plasma conditions, (b) the characteristics of the interactions between these instability modes, their transition dynamics and role toward the inverse energy cascade and self-organization phenomena [34], and (c) the contributions of these instabilities to the particles' and energy transport across the magnetic field and the consequent influence on the global plasma behavior.

With these main questions in mind, a relevant and recent prior computational work to this effort has been reported in Ref. [35], in which the author carried out some parametric studies on the MTSI and the ECDI instabilities in a radial-azimuthal setup that did not resolve the plasma sheath. In that work, the instabilities' characteristics for two values of the axial electric field was studied for the xenon propellant [35]. The author also proposed a simplified expression for the dependency of the azimuthal wavenumber of the MTSI's fastest growing mode to the electric and magnetic fields [35], which we will discuss in more detail in Section 3.2. Another recent research of similar scope to the present paper was carried out by Croes et al [36]. In that work, for a single value pair of the electric and magnetic field intensities, the authors assessed the characteristics of the ECDI and the electron transport due to this instability in a collisional radial-azimuthal simulation setup featuring secondary electron emission and with xenon, krypton, argon, and helium propellants. Most notably, they observed that, in line with the existing theories, the frequency of the ECDI depends on the ion mass [36]. However, they did not notice a notable variation in the electron transport with the choice of the propellant [36].

Our research reported here delves deeper into the subjects examined by these previous efforts. In particular, we assess across several values of the electric field intensity the variations in the development and the characteristics of the ECDI as well as the MTSI modes in the presence of the plasma sheath. The more controlled setup of our simulations, which excludes the effects of collisions and the secondary electron emission in contrast to Ref. [36], enables identifying the direct influences of the $E$-field's intensity. Finally, we have adopted novel spectral analysis technique that provides unprecedented insights into the spatial structures of the instability modes.

### Section 2: Description of the simulations' setup and conditions

The simulations' setup in this study follows, in general, the 2D radial-azimuthal benchmark case, which was introduced in Ref. [33], and that we had adopted in previous publications to verify the reduced-order PIC scheme [16] and to carry out parametric physics studies [17].

The setup is representative of a radial-azimuthal cross-section of a Hall thruster and features an externally applied axial electric field ($E_y$) and radial magnetic field ($B_x$). Regarding the notation of the axes, $x$, $y$, and $z$ represent, respectively, the radial, axial, and azimuthal directions.

The simulations are performed with three different ion species (propellants), namely, xenon (Xe), krypton (Kr), and argon (Ar). Moreover, in this part I of the article, the $E_y$ magnitude is varied to investigate the impacts on the involved physics. The values of the electric field intensity in the simulated cases are presented in Table 1.

| Case No. | $E_y\ [kVm^{-1}]$ | $E_y/E_0$ | Propellant |
|---|---|---|---|
| 1 | 5 | 0.5 | Xe, Kr |
| 2 | 10 | 1 | Xe, Kr, Ar |
| 3 | 20 | 2 | Xe, Kr, Ar |
| 4 | 30 | 3 | Xe, Kr, Ar |
| 5 | 40 | 4 | Xe, Kr, Ar |
| 6 | 50 | 5 | Xe, Kr, Ar |

Table 1: List of the studied simulation cases, summarizing the value of the axial electric field ($E_y$) and the ion species used in each case.

The electric field of $1E_0$ ($10\ kVm^{-1}$) refers to the baseline value in the benchmark setup [33]. In all simulations reported in this part I, $B_x$ is $20\ mT$, which is the same value as that in the benchmark setting [33]. The remaining conditions and setup details, which are shared among all simulations, are described in the following.

The simulation domain is a 2D Cartesian plane with equal extent of 1.28 cm along both the radial and the azimuthal directions. The cell size is considered to be $50\ \mu m$ corresponding to 256 nodes along either dimension. The total simulated time is $30\ \mu s$ with a time step of $1.5 \times 10^{-11}\ s$.

Initially, electrons and ions are sampled from Maxwellian distribution functions at the temperatures of 10 eV and 0.5 eV for the electrons and ions, respectively. The particles are loaded on the simulation plane with a uniform distribution corresponding to the density of $1.5 \times 10^{16}\ m^{-3}$. The initial number of macroparticles per cell is 100.

The simulations are collisionless and, to compensate the fluxes of particles to the wall and, hence, to achieve a steady-state density in the simulations, a particle injection source is imposed. This source has an azimuthally uniform distribution and a cosine profile along the radial direction extending from $x = 0.09$ cm to $x = 1.19$ cm. The peak value of the injection source along the $x$ direction is $8.9 \times 10^{22}\ m^{-3}s^{-1}$. At each time step, electron-ion pairs are injected according to the radial profile of the injection source with velocities sampled from Maxwellian distribution functions at the initial temperatures of the respective species.

As for the boundary conditions, in terms of the electric potential, a zero-volt Dirichlet condition along the radial and a periodic condition along the azimuthal direction is applied. With respect to the particles, all particles reaching the walls are removed and no secondary electron emission is considered. Particles leaving the domain along the azimuth re-enter from the other azimuthal end with the same velocity and radial position to mimic periodicity condition. As the simulations do not resolve the axial direction, in order to allow the particles' energy to reach a steady-state [37], a finite artificial extent of 1 cm is assumed along the $y$ direction on both sides of the radial-azimuthal simulation plane [16][33]. All particles crossing the axial boundaries are resampled from the initial Maxwellian distributions and reloaded on the simulation plane while maintaining their radial and azimuthal positions.

All simulations are performed using the IPPL[1]-Q2D code. The domain decomposition associated with the reduced-order PIC scheme is applied along the radial and azimuthal dimensions of the simulations using 50 regions [16].

**Section 3: Results and discussion**

In this section, we present and discuss the impact(s) of the magnitude of the imposed axial electric field on various aspects of the plasma dynamics. The results are presented for the three named propellants, Xe, Kr and Ar, and the notable differences observed in the discharge behavior among these propellants are emphasized. It is important to highlight that, as the simulations are collisionless, any variation observed in the simulations with different propellants is solely due to the difference in the ions' mass.

The results are presented in terms of the influence of the $E_y$ intensity on the plasma properties' profiles as well as the dominant instability modes and their implications on the induced axial electron transport and particles' distribution functions. In the case of xenon propellant, the instabilities' contribution to the electron transport for each $E_y$ value is compared against the contribution of other force terms that appear in the electron azimuthal momentum equation.

In addition, to isolate the individual spatiotemporal coherent structures present in each simulation case, the results of the Dynamic Mode Decomposition (DMD) analysis [38] are provided. Finally, the azimuthal wavenumbers of the dominant instabilities for various values of $E_y$ are compared against their respective theoretical values to evaluate the degree of consistency between the simulations' results and the available theories.

**3.1. Variation in the plasma properties distribution**

We first present the time-averaged radial profiles of the relevant plasma properties, which are averaged over 20-30 $\mu s$. This averaging window corresponds to when the simulations have reached quasi-steady state. Figure 1(a) shows the radial distributions of the ion number density ($n_i$) for different propellants and with various $E_y$ intensities. The change in the peak value of the number density profiles vs $E_y$ are plotted in Figure 1(b).

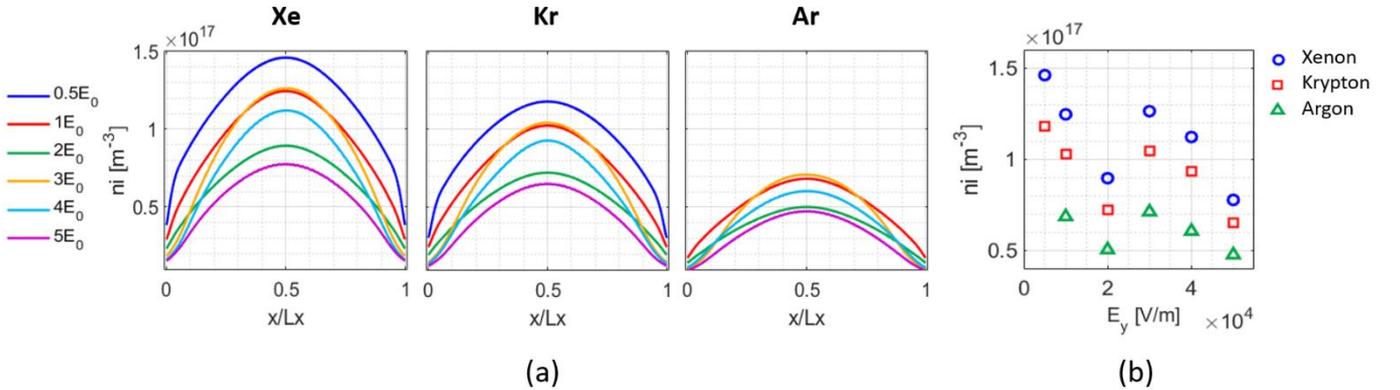

Figure 1: (a) Radial profiles of the ion number density ($n_i$) averaged over 20-30 $\mu s$ from the radial-azimuthal simulations with various axial electric field intensities and propellants. (b) Variation vs $E_y$ of the ion number density's peak value along the radial direction for the three studied propellants.

From these plots, it is evident that the steady-state density established in the simulation is the lowest for Ar and the highest for Xe across various $E_y$ values. However, the $n_i$ peak and the overall radial profiles follow the same general trend for the three propellants when varying $E_y$. The $n_i$ peak decreases as $E_y$ increases from 5-20 $kVm^{-1}$, then, after experiencing a jump in values between 20-30 $kVm^{-1}$, it continues its decreasing trend beyond $E_y = 30$ $kVm^{-1}$.

Such non-monotonic behavior is also observed in Figure 2(c) and (f) for the radial electron temperature ($T_{ex}$), where $T_{ex}$ is seen to be plateaued for the $E_y$ range of 20-30 $kVm^{-1}$. This is whereas the total electron temperature ($T_e$) and the azimuthal electron temperature ($T_{ez}$) monotonically increase with $E_y$ as it can be seen from Figure 2(a) and (d) and Figure 2(b) and (e), respectively.

---

[1] Imperial Plasma Propulsion Laboratory

These observed trends from Figure 1 and Figure 2 can be explained by considering two facts. First, it is noted that prescribing a fixed injection source creates a coupling between the radial electron temperature and the number density within the simulation [16][33]. if $T_{ex}$ increases, the flux of particles to the wall increases and, since the rate of particles' injection is not dependent on the electron temperature (as opposed to a self-consistent ionization case, for example), the source cannot fully compensate the particle losses to the wall. This leads to a reduction in the number density. Similarly, in case the $T_{ex}$ drops, the source overcompensates the particles loss to the wall and, hence, the density increases.

Second, it has been demonstrated that the ECDI has a significant role in the azimuthal heating of the electrons through particle-wave interactions along the azimuthal direction [32][39], whereas the MTSI is observed to heat up the electrons mainly in the radial direction parallel to magnetic field [24][40]. Thus, the deviation from a monotonic increase in $T_{ex}$ with $E_y$ after 30 $kVm^{-1}$ (Figure 2(f)) roots in the existence of a weakened MTSI, which have resulted in a lower radial heating than expected compared to a case in which the variation with $E_y$ would have been fully monotonic. This explanation is further evidenced in the next sections through the Fast Fourier Transform (FFT) and detailed DMD analyses of the instabilities. In any case, according to the point in the preceding paragraph, if $T_{ex}$ deviates from a monotonic increase with $E_y$, this will be reflected in the variation with $E_y$ of the $n_i$ in terms of a non-monotonic decrease (Figure 1(b)).

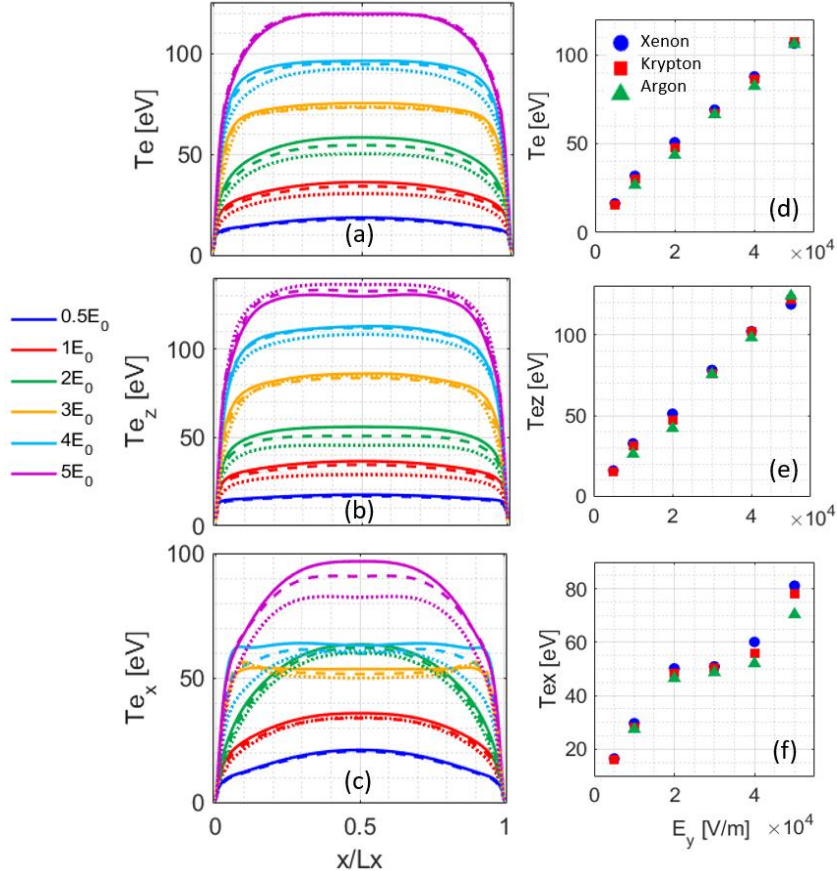

Figure 2: (Left Column) Radial profiles of (a) electron temperature ($T_e$), (b) azimuthal electron temperature ($T_{ez}$), and (c) radial electron temperature ($T_{ex}$), averaged over 20-30 $\mu s$ from the radial-azimuthal simulations with various axial electric field intensities and propellants; solid lines correspond to xenon, dashed lines to krypton, and dotted lines to argon. (Right Column) Variation vs $E_y$ of the radially averaged $T_e$ (plot (d)), $T_{ez}$ (plot (e)), and $T_{ex}$ (plot (f)) for the three studied propellants from the quasi-2D simulations.

In order to identify with further accuracy the value of $E_y$ after which MTSI become weaker, thus, the variation of with $E_y$ of the $n_i$ and $T_{ex}$ departs from monotonic trend, additional simulations were performed with Xe to cover the $E_y$ range of 5 to 50 $kVm^{-1}$ with a smaller increment (every 2.5 $kVm^{-1}$). The resulting variation of the ion number density's peak and the radial electron temperature's mean with $E_y$ are shown in Figure 3. From these plots, we notice that it is around the axial electric field strength of $25\ kVm^{-1}$ where the change in the behavior occurs. This critical $E_y$ value seems to be extendable to other propellants as well based on the similarity of trends among various gases as seen in Figure 1(b).

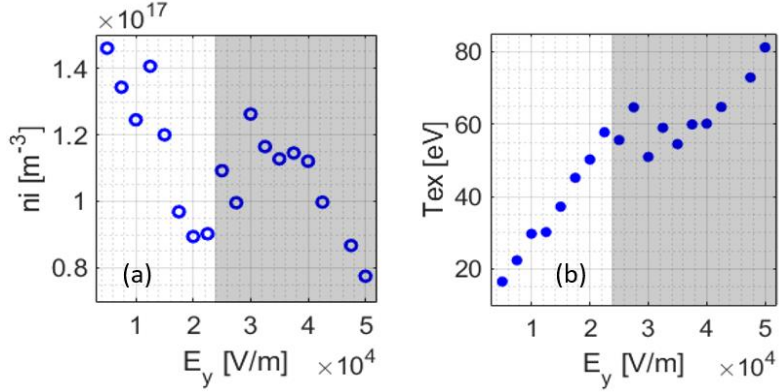

Figure 3: Variation vs $E_y$ values with $0.25E_0$ increments of the peak ion number density (a) and radially averaged radial electron temperature (b) for the quasi-2D simulations with xenon propellant. The shaded range of $E_y$ values correspond to a change in the variation trend of the plasma properties compared to the smaller values of $E_y$.

In the following, we demonstrate that the simulations with $E_y$ values below and above this "threshold" exhibit two distinct general behaviors due to different relative strengths of the MTSI and the ECDI within the two different $E_y$ ranges.

Considering that the effect of the ECDI and the MTSI on electrons' heating are mainly along the azimuthal and the radial directions, respectively, the ratio of the radial-to-azimuthal electron temperature ($T_{ex}/T_{ez}$), which is a measure of anisotropy, can be an indicator of the relative significance of these two instability modes. In this respect, the radial profiles of the electron temperature ratio and their mean values throughout the domain vs the $E_y$ value and for the different propellants are presented in Figure 4 (top row).

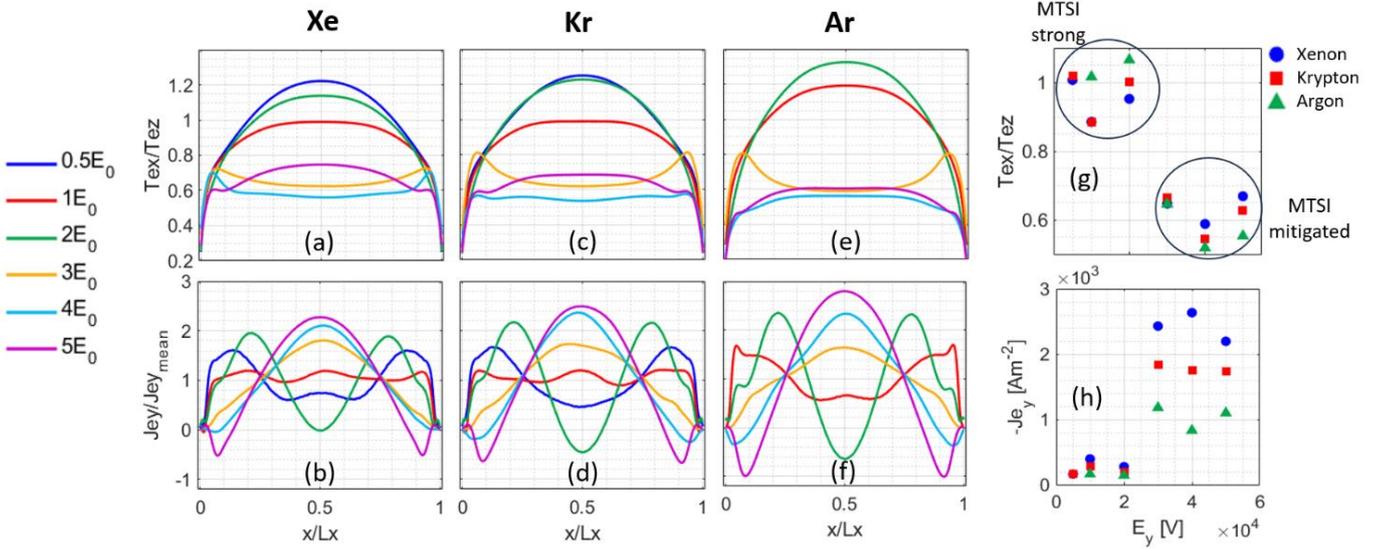

Figure 4: (a)-(f) Radial profiles of the radial-to-azimuthal electron temperature ratio ($T_{ex}/T_{ez}$) (top row) and the normalized axial electron current density ($J_{ey}/J_{ey,mean}$) (bottom row) averaged over 20-30 $\mu s$ from the quasi-2D simulations with various axial electric fields and propellants. (g) Variation vs $E_y$ of the radially averaged values of $T_{ex}/T_{ez}$ for the quasi-2D simulations with various propellants. (h) Variation vs $E_y$ of the radially averaged values of $-J_{ey}$ for the quasi-2D simulations with various propellants.

The mean values in plot (g) show two distinct clusters which indicate two different operating regimes: regime I, where $\frac{T_{ex}}{T_{ez}} \approx 1$ for $E_y$ between 5-20 $kVm^{-1}$, and, regime II, where $\frac{T_{ex}}{T_{ez}} < 1$ for $E_y$ between 30-50 $kVm^{-1}$. Regime I is characterized by the presence of strong MTSI modes with a radial heating that is almost comparable with or larger than the azimuthal heating of the ECDI. Whereas, in regime II, the MTSI is mitigated, which leaves the ECDI as the dominant instability mode, thus, leading to a larger $T_{ez}$ than $T_{ex}$. From Figure 4(g), the temperature ratios are overall seen to be the highest for Ar and the lowest for Xe in regime I whereas the opposite is true in regime II. This means that Ar exhibits a larger anisotropy in the electron temperature.

Regarding the radial distribution of $T_{ex}/T_{ez}$ (plots (a), (c) and (e) in Figure 4,), various cases exhibit different profiles. In simulations with the axial electric field of 5 and 20 $kVm^{-1}$ ($0.5E_0$ and $2E_0$ cases), the temperature ratio is large at the center of the domain and become smaller away from the centerline. The variation of the temperature ratio from the centerline toward the walls is more moderate in the cases with the $E_y$ strength of 1 $kVm^{-1}$ ($1E_0$ case). However, for $E_y = 30$ $kVm^{-1}$ ($3E_0$ case), the temperature ratio is larger near the wall than in the central region. At the field intensity of 40 $kVm^{-1}$ ($4E_0$ case), each propellant shows different behavior. With Xe, $T_{ex}/T_{ez}$ ratio is larger close to the walls, whereas it is higher near the center for Ar, and, in the case of Kr, the profile is rather radially uniform. Finally, at 50 $kVm^{-1}$ ($5E_0$ case), all propellants indicate higher temperature ratios at the central part of the domain.

The bottom row of Figure 4 illustrates the radial profiles of the normalized axial electron current density ($J_{ey}/J_{ey,mean}$) for different simulation cases (plots (b), (d), and (f)), together with the variation vs $E_y$ of the absolute value of the mean axial electron current density (plot (h)). It is evident from these plots that the $J_{ey}$ has different distributions in various cases. For cases where the MTSI is the dominant mode, including $0.5E_0$ and $2E_0$, $J_{ey}$ is mostly largest away from the center. This is sort of a characteristic profile of the electron transport induced by the MTSI, which we also observed in Ref. [17] for the conditions where the MTSI appears. In contrast, for the cases with larger axial electric field ($3E_0$, $4E_0$ and $5E_0$), in which ECDI is dominant, the axial electron current is mostly concentrated toward the center and, in $4E_0$ and $5E_0$ cases, reverse its direction close to the walls. For the baseline condition ($1E_0$) and for Xe and Kr, the ECDI and the MTSI waves are of comparable amplitudes and their combined contributions to the electron transport has led to a more uniform current distribution across the domain's radial extent. However, the same baseline conditions for Ar results in the MTSI to develop with a larger amplitude relative to the ECDI. This causes $J_{ey}$ to be less uniform and have peaks toward the walls.

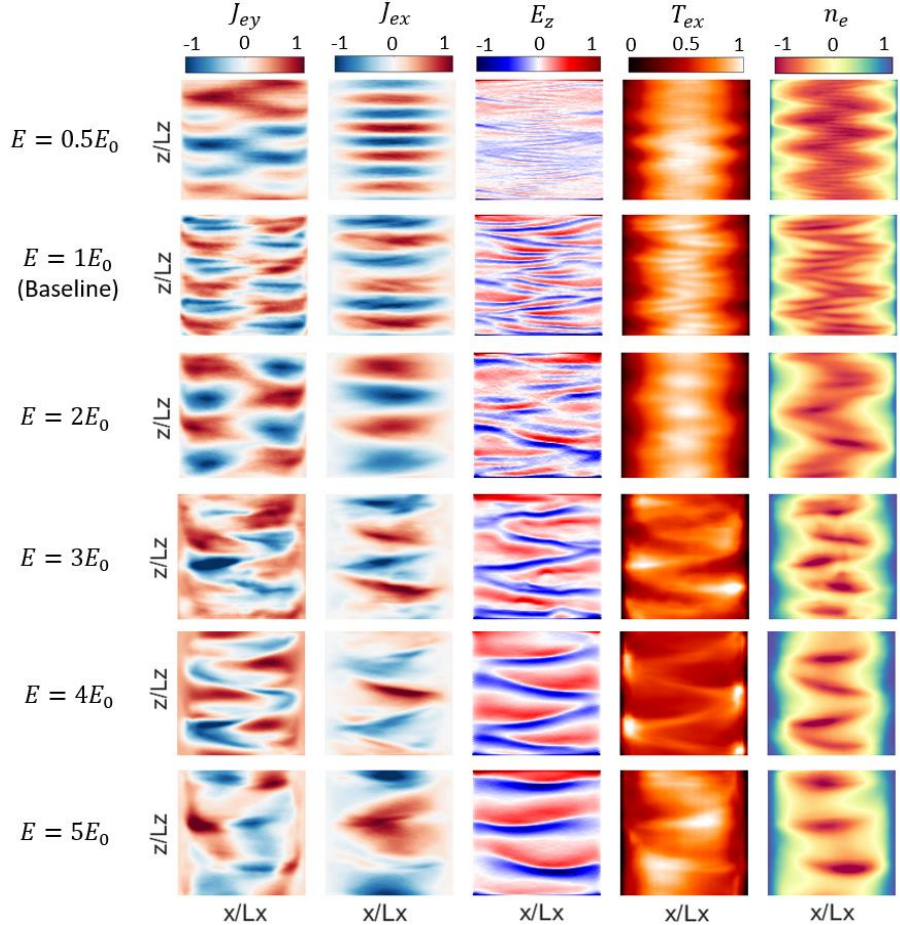

Figure 5: Comparison of the 2D snapshots of the normalized plasma properties at a time of local maximum of the radial electron temperature for various axial $E$-field intensities and the xenon propellant. The columns, from left to right, represent

the axial and radial electron current densities ($J_{ey}$ and $J_{ex}$), the azimuthal electric field ($E_z$), the radial electron temperature ($T_{ex}$), and electron number density ($n_e$).

An additional interesting observation is that the variation of $J_{ey}/J_{ey,mean}$ across the radial extent of the domain is the largest for Ar and the smallest for Xe, whereas the mean magnitude of the current density across the domain increases from Ar to Xe for all $E_y$ cases.

Before presenting the FFT spectra and the DMD modes of the azimuthal electric field from various simulation cases to quantitatively assess the characteristics of the involved ECDI and MTSI modes, looking at the snapshots of the plasma properties can provide a qualitative comparison of the dominant instabilities in the simulated cases. In this regard, Figure 5 illustrates the 2D snapshots of various plasma properties for Xe at a sample instant of time when the radial electron temperature is maximum. This corresponds to moments along the discharge evolution that the MTSI, if excited in the simulation, is fully developed with a high amplitude. The MTSI patterns (which are characterized by both azimuthal and radial wavevector components) are easily recognizable in the snapshots of the cases with $0.5E_0$, $1E_0$ and $2E_0$. In the remaining cases, the MTSI pattern gets distorted in the $J_{ey}$ snapshots. However, the fluctuations still show a finite wavenumber in the radial direction except for $5E_0$ case, in which the fluctuations become mostly azimuthal. The azimuthal wavelength of the waves becomes larger with increasing $E_y$, which we will demonstrate later to be consistent with the theory.

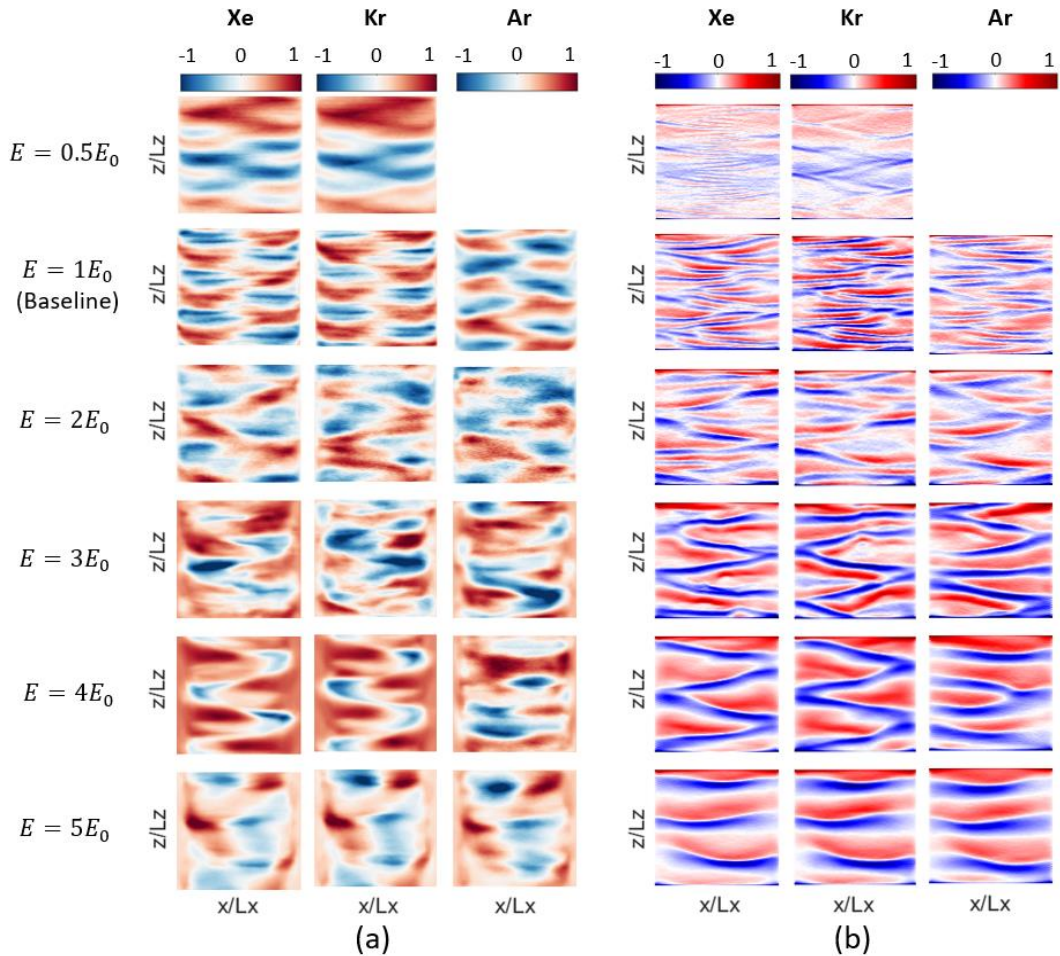

Figure 6: Comparison of the 2D snapshots of (a) normalized radial electron current density ($J_{ey}$), and (b) the normalized azimuthal electric field ($E_z$) at a time of local maximum of the radial electron temperature for various axial $E$-field intensities and the three propellants.

Figure 6 compares the 2D snapshots of the axial electron current density and the azimuthal electric field between Xe, Kr, and Ar for various $E_y$ cases. From this figure, it is evident that the snapshots are rather similar in most cases across all propellants. The cases with noticeable difference between the three propellants include the $0.5E_0$ simulation, where the ECDI waves with short azimuthal wavelength are absent for Kr. Also, in the $1E_0$ case (baseline) with Ar, the azimuthal ECDI waves are less apparent in the $J_{ey}$ snapshot compared to the other propellants. This suggests a mitigated ECDI with a lower amplitude in this case as was hinted at within the previous discussions.

### 3.2. Variation in the characteristics and structure of azimuthal instabilities

To evaluate the wavenumber and frequency of the wave modes, we start by presenting in Figure 7 the spatiotemporally averaged spatial FFT of the azimuthal electric field ($E_z$). The average is taken over FFT at all radial positions and over the time interval of 20-30 $\mu$s. In the plots of Figure 7, the horizontal axis is normalized by the fundamental resonance wavenumber of the ECDI ($k_0 = \frac{\Omega_{ce}}{V_{d_e}}$, where $\Omega_{ce}$ and $V_{d_e}$ are the electron cyclotron frequency and the electron azimuthal drift velocity, respectively).

The FFT spectra of cases with lower $E$-field ($0.5E_0, 1E_0$ and $2E_0$) contain distinct peaks at $k_z/k_0 \approx 0.2$ corresponding to the MTSI wave modes. All FFT spectra additionally show the presence of the first harmonic of the ECDI ($k_z/k_0 \approx 1$) except for the $0.5E_0$ case with Kr. Furthermore, whereas in most cases, the second harmonic of the ECDI ($k_z/k_0 \approx 2$) is also visible, the FFT spectrum of $5E_0$ case have peaks at the third and fourth ECDI harmonics (not specified in the figure) as well for all propellants.

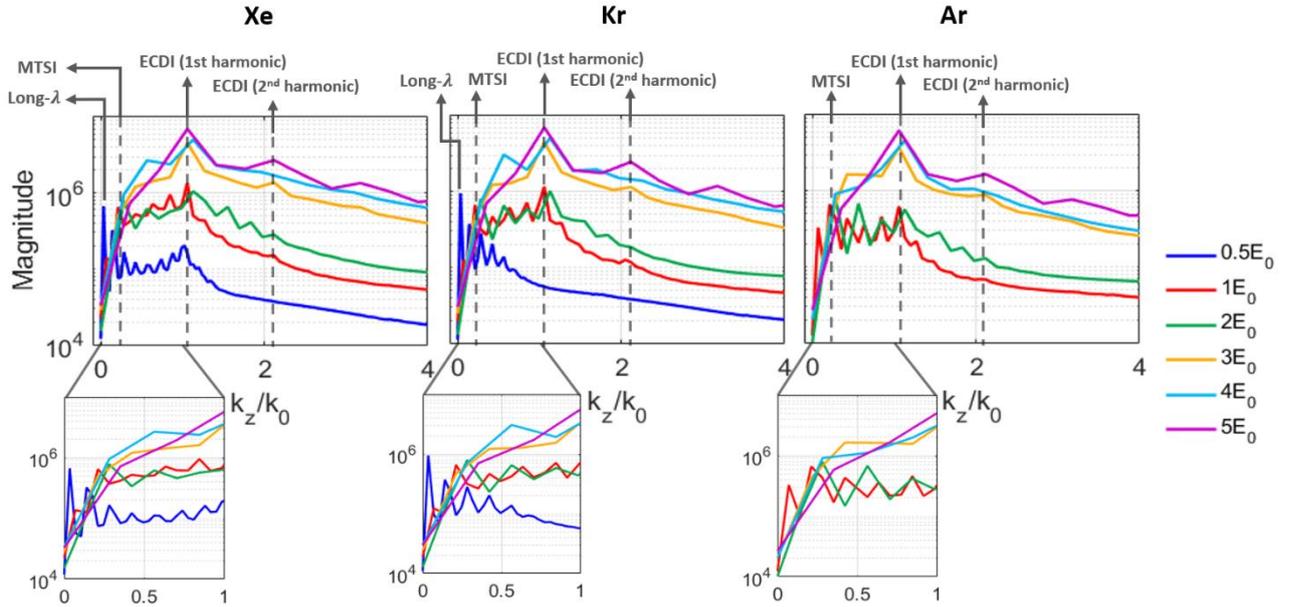

Figure 7: 1D spatial FFT plots of the azimuthal electric field signal from simulations with various values of $E_y$ and different propellants. The FFTs are averaged over all radial positions and over the time interval of 20-30 $\mu s$.

The spatially averaged temporal FFT of $E_z$ for different values of $E_y$ and various propellants are provided in Figure 8. It is noticed regarding this figure that, in some cases, the temporal FFT has many peaks, which makes it difficult to identify the constituent modes and find the frequencies associated with the peaks found in the spatial FFT.

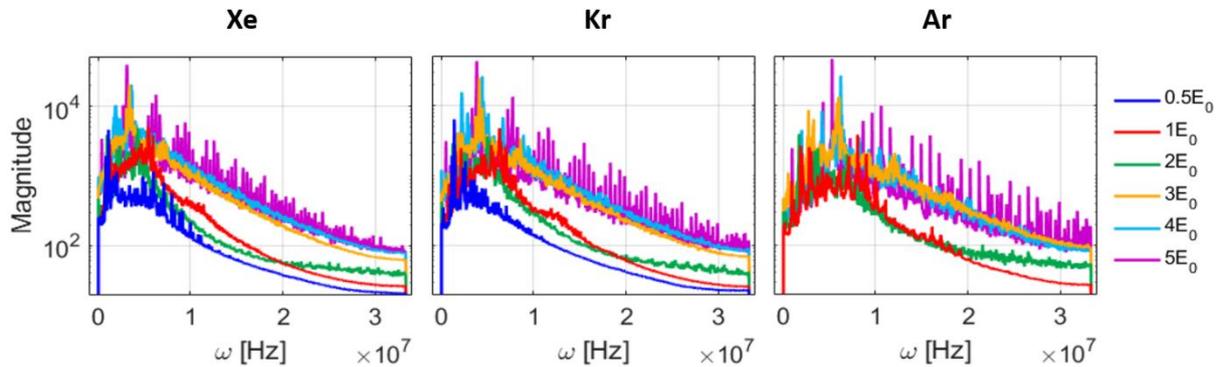

Figure 8: 1D spatially averaged temporal FFT plots of the azimuthal electric field signal from quasi-2D simulations with various magnitudes of the axial electric field and different propellants.

For example, in simulation cases with $3E_0, 4E_0$ and $5E_0$, although the highest peaks with the frequency ranges of about $3-3.5\ MHz$ for Xe, $4-4.5\ MHz$ for Kr and $5.5-6\ MHz$ for Ar corresponding to the dominant ECDI wave modes are readily distinguishable, at lower frequencies close to the MTSI frequency in each case, there are several peaks in close proximity to one another. In addition, compared to the simulations with Kr and Xe, the FFTs for Ar are more crowded by peaks of similar amplitude (see, for instance, the $1E_0$ case). It is also noticed that, in the $5E_0$ simulation case, the FFT spectra contain multiple clusters of peaks for all propellants, in each of which, the highest peak represents a harmonic of ECDI. In this respect, from the plots in Figure 8, we may distinguish 5 to 6 ECDI harmonics.

As a result of the above challenges toward characterizing the instability modes using FFT, we have employed the optimized DMD (OPT-DMD) method [41] to analyze more effectively and in greater detail the instabilities' spectral characteristics. In particular, we would like to find out what physical instability modes the peaks in the FFT spectra correspond to and how the spatial structures associated with each of these modes look like.

The central concept of the DMD lies in the decomposition of a time-series data into a set of spatial patterns, known as modes, and their associated temporal evolution. As the underlying assumption of the DMD is the linearity of the time dynamics, the temporal evolution of each mode is represented by a complex frequency, i.e., it can have both an imaginary part for purely oscillatory behavior as well as a real part to account for growth or damping. A great advantage of the DMD over the FFT analysis is that contrary to spatial FFT which assumes purely harmonic bases in space, the DMD spatial bases are arbitrary and specific to the data it is applied to. As a result, the DMD calculates the spatial structure of the instability modes within the entire domain from which information such as the wavelength of the instabilities along all simulation dimensions can be simultaneously inferred.

Noting in addition the OPT-DMD's stability and robustness [38][41], this approach is a powerful tool to analyze complex dynamical systems by extracting and isolating the dominant coherent modes from the time-series snapshots of the system. A detailed description of the general DMD method can be found in Ref. [38], where we demonstrated as well the OPT-DMD's application to analyze the PIC simulations' data across multiple test cases particularly for the identification and isolation of the spatiotemporal behavior of the involved instabilities.

The OPT-DMD is applied in this work to the time series snapshots of the azimuthal electric field in the time interval of 20-30 $\mu s$, with each snapshot being $1.5 \times 10^{-2} \mu s$ apart. The derived $E_z$ spatial modes together with their associated frequencies are presented in Figure 9 for Xe and for various axial $E$-field intensities. The modes shown in this figure are those whose frequency corresponds to the distinct peaks in the temporal FFT (Figure 8). In addition, the modes delimited by the dark blue and red boxes refer, respectively, to the wave modes whose wavelength matches the wavelengths detectable for the MTSI and ECDI from the spatial FFT (Figure 7). The outlined MTSI modes are characterized by clear radial-azimuthal structures, though their exact patterns differ from one case to another, and their azimuthal wavelengths become longer as $E_y$ increases. The outlined ECDI modes exhibit purely azimuthal structures with almost uniform amplitudes across the domain for all cases except for $E_y = 0.5E_0$ and $1E_0$ for which the ECDI amplitude is larger in the central part of the domain. In the $0.5E_0$ case, a long-wavelength mode has additionally appeared with a wavelength almost the same as the domain's azimuthal extent.

Apart from the modes which we have identified as the ECDI and the MTSI, there are other modes exhibiting various spatial structures. For low $E$-field cases ($E_y \leq 20\ kVm^{-1}$), several modes have a radial wavenumber, whereas, for higher $E$-field values, most modes are almost purely azimuthal. These modes whose spatial patterns do not exactly match the well-known ECDI or MTSI instabilities could represent the intermediate states during the modes' transition. In fact, the nonlinear interactions between the instabilities that exist in the adopted radial-azimuthal simulation configuration [33] can lead to the periodic growth and damping of various instabilities in the system. Therefore, the DMD-identified modes other than the labelled ones in Figure 9 may represent the variation in the instabilities' spatial structure during the discharge evolution as they transition from one to another.

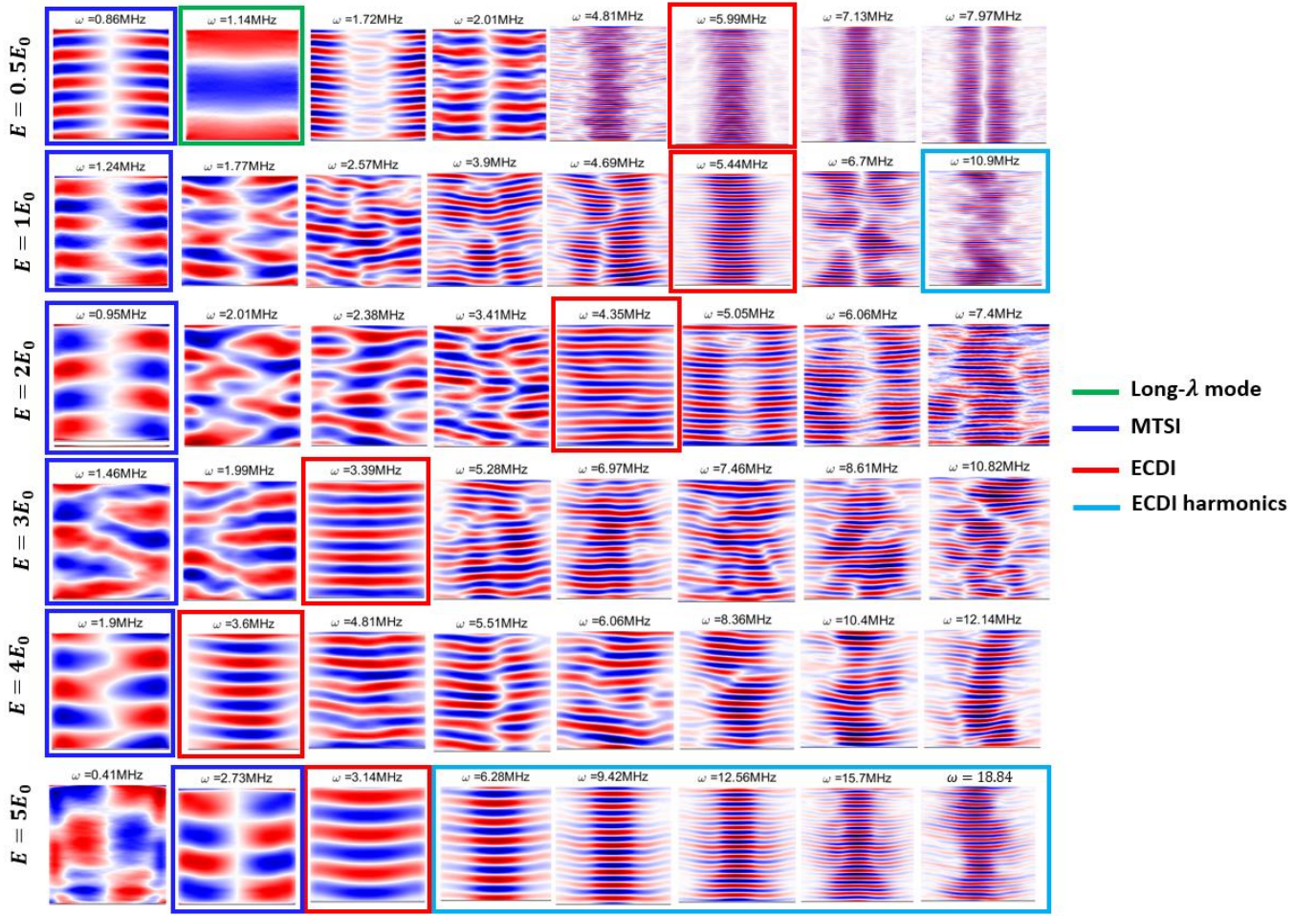

Figure 9: Visualization of the several dominant DMD modes of the azimuthal electric field data from the radial-azimuthal simulations with different $E_y$ values and the xenon propellant. The approach pursued to derive these modes is explained in detail in Ref. [38]. The modes identified with a box correspond to those with wavelengths equal to the wavelengths of the dominant modes visible in the spatial FFT plot in Figure 7.

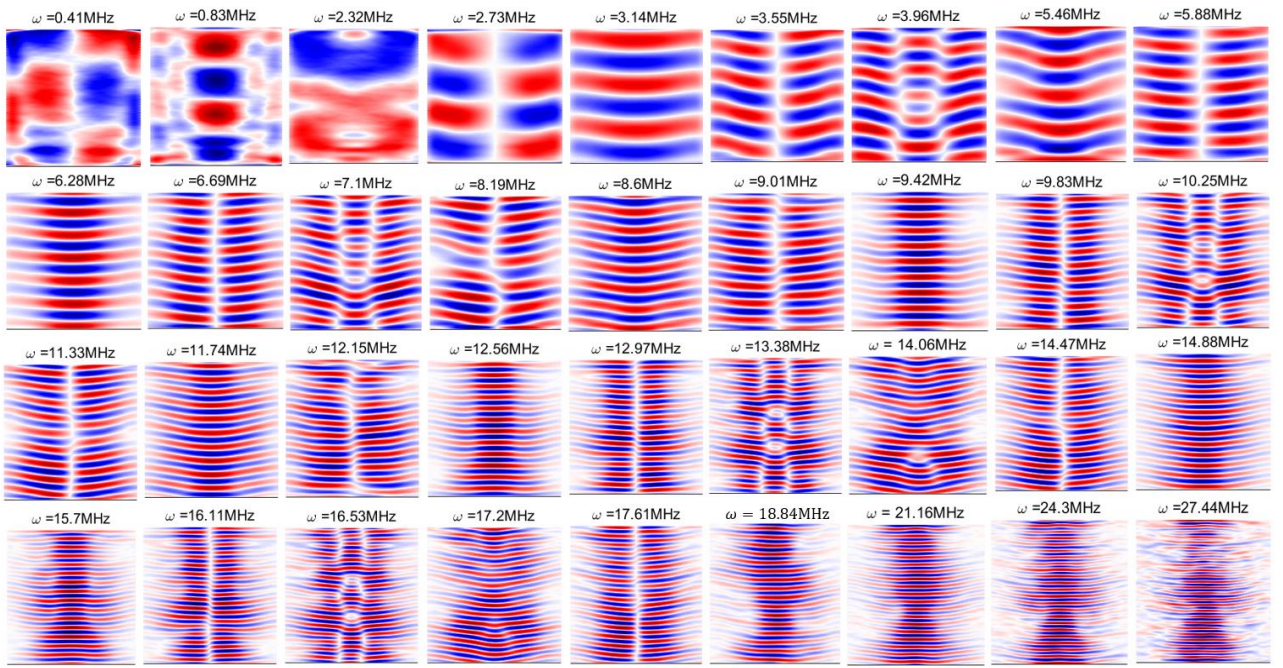

Figure 10: Visualization of the expanded set of consecutive dominant DMD modes of the azimuthal electric field data from the radial-azimuthal simulation with $E_y = 5E_0$ and the xenon propellant.

For the $E_y = 5E_0$ case with xenon propellant, five ECDI harmonics are captured in the simulation and, thus, an expanded set of DMD modes associated with the many peaks observed in the corresponding temporal FFT plot (Figure 8, left) are shown in Figure 10. In terms of the spatial structures of these modes, we can observe three distinct patterns which are repeated at various frequencies and wavelengths. These patterns include purely azimuthal oscillations, azimuthal-radial patterns with the radial wavelength matching the domain's extent (like the MTSI), and an azimuthal structure which is distorted near the center of the domain.

In Figure 11, we can see the comparison of the spatial modes corresponding to the MTSI and the ECDI for different propellants and at various $E$-field values. Although the overall patterns of these modes are consistent among Xe, Kr and Ar simulations, slight differences are visible in some cases. For instance, for $3E_0$ and $4E_0$ conditions, the MTSI modes exhibit rather different patterns in Ar compared to Xe.

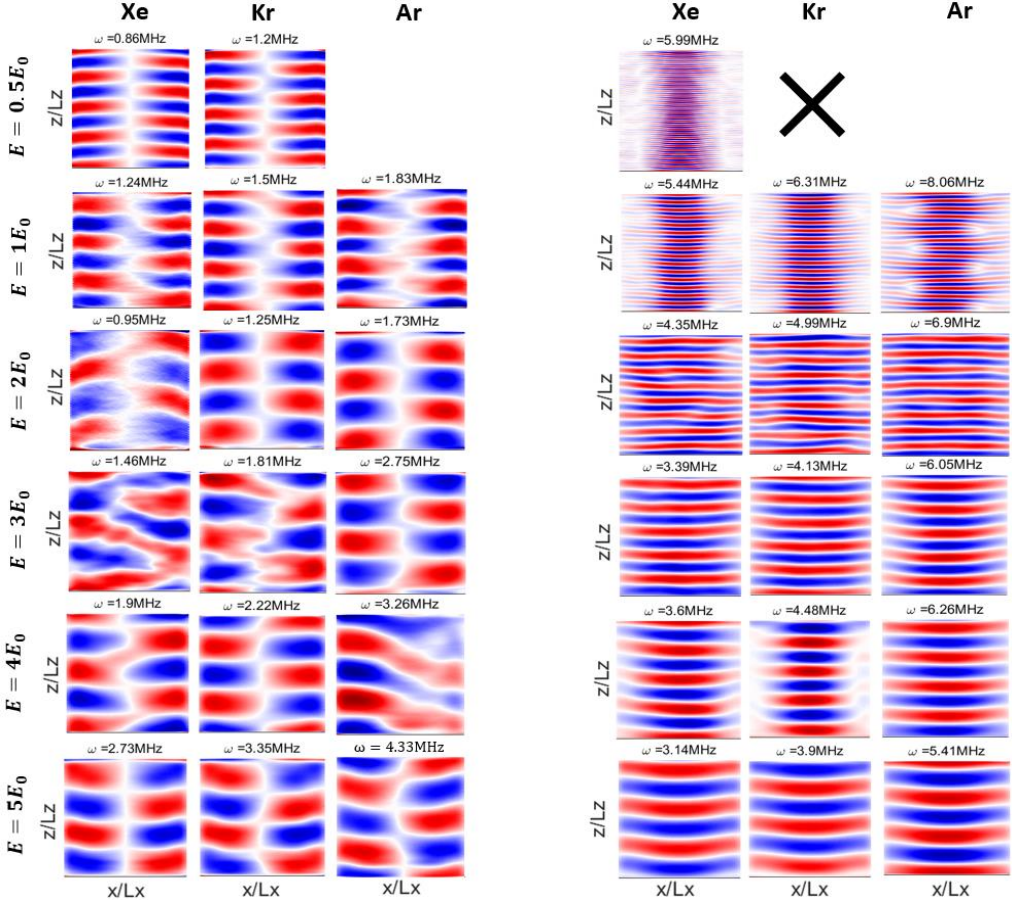

Figure 11: Visualization of the dominant MTSI (left) and ECDI (right) modes from the DMD of the azimuthal electric field data from the radial-azimuthal simulations with different $E_y$ values and the three propellants.

To conclude the assessments of the instabilities' characteristics, the azimuthal wavenumber, the frequency, and the phase velocity of the MTSI and the first harmonic of the ECDI for different propellants are plotted against $E_y$ in Figure 12. The wavenumbers of these two instabilities from the simulations are compared against the analytical values derived from their respective dispersion relations.

Concerning the ECDI, the resonance condition occurs near certain azimuthal wavenumbers as given by Eq.1 [2][33]. In this equation, $k_z$ is the azimuthal wavenumber, $e$ is the elementary charge, and $m_e$ is the electron mass. $k_z$ is noticed from Eq. 1 to be independent of the ion mass (propellant type) and varies proportionally with $B_x^2$ and inversely with $E_y$. From Figure 12 (top row, left panel), the variation of the ECDI's azimuthal wavenumber with $E_y$ from the simulations is seen to be consistent with the theoretical variation according to Eq. 1.

$$k_z = n \frac{\Omega_{ce}}{V_{d_e}} = n \frac{e}{m_e} \frac{B_x^2}{E_y} \ ; \ n = 1, 2, \ldots \qquad \text{(Eq. 1)}$$

Regarding the MTSI, Ref. [35] recently proposed a relationship (Eq.2) between the radial and azimuthal wavenumbers of the fastest growing mode of the MTSI. The relationship indicates the resonance condition for the MTSI, which is derived from the radial-azimuthal fluid dispersion relation of the instability with cold electrons and after making simplifying assumptions. The details of the derivation of the MTSI's dispersion relation can be found in Ref. [35].

$$k_z = \sqrt{\frac{e}{m_e}\frac{B_x^2}{E_y}k_x} \quad \text{(Eq. 2)}$$

In Eq.2, $k_x$ is the radial wavenumber of the instability. From the results presented above, the radial wavelength of the MTSI excited in our simulations is twice the radial size of the domain, meaning that $k_x \approx 490\ rad/m$ for our cases.

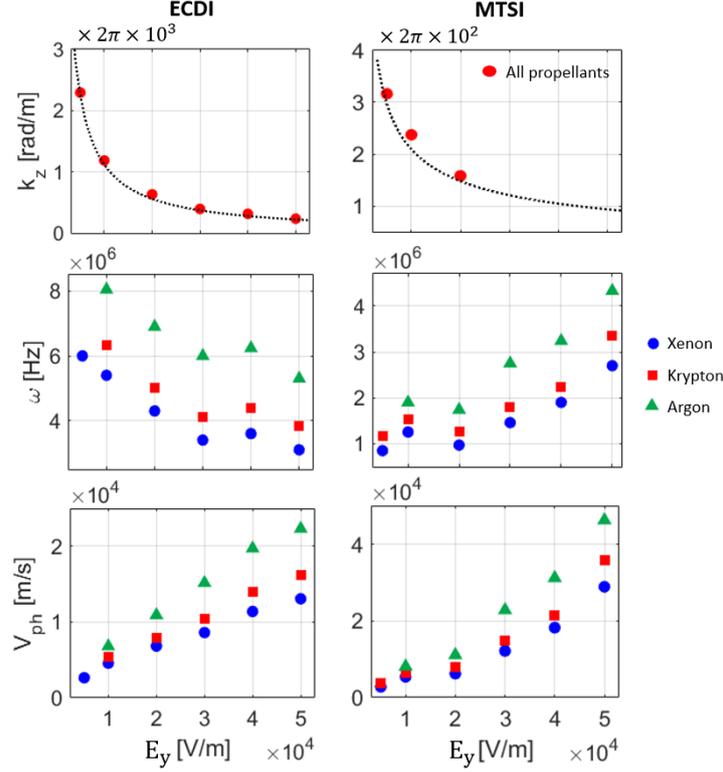

Figure 12: Variation vs $E_y$ of the characteristics of the dominant ECDI and MTSI modes from the radial-azimuthal simulations with different propellants. (Top row) variation of the azimuthal wavenumber ($k_z$) compared against the theoretical relations (Eqs. 1 and 2) for the wavenumber of ECDI's first harmonic and the MTSI's fastest growing mode shown by black dotted lines, (middle row) variation of the real frequency ($\omega$), (bottom row) variation of the azimuthal phase velocity ($V_{ph}$).

Despite the simplifications made to derive the analytical relation in Eq. 2, Ref. [35] verified its agreement in predicting the location of the fastest growing mode in the $k_x - k_z$ plane with that obtained from the kinetic dispersion relation. This relationship also denotes a threshold that delimits the stable and unstable conditions for the MTSI in a simulation [35]. Based on this stability criterion, the MTSI is unstable if $k_z \lesssim \sqrt{\frac{e}{m_e}\frac{B_x^2}{E_y}k_x}$.

According to Eq.2, the azimuthal wavenumber of the MTSI does not depend on the ion mass; it increases proportionally with $B_x$ and decreases with the $\sqrt{E_y}$ factor. This is observed from Figure 12 (top row, right panel) to be in line with what is predicted by our simulations.

It is noted that the plotted azimuthal wavenumbers of the MTSI and the ECDI from the simulations correspond to the respective peaks in the spatial FFTs (Figure 7). However, the plot only includes the MTSI wavenumbers for the low $E$-field cases ($0.5E_0$, $1E_0$, and $2E_0$). This is because, even though the MTSI modes are identified for all cases from the DMD analysis, for cases with $E_y$ between $30 - 50\ kVm^{-1}$, clear peaks for the MTSI are not detected in the spatial FFT spectra. The reason for this discrepancy is that the MTSI in these cases is not strong enough at all times to be clearly visible in the time-averaged FFT spectrum. Nonetheless, by appropriately choosing the truncation rank for the DMD [38], this method is capable of identifying the relatively low-energy MTSI modes as well.

Regarding the frequencies of the instabilities, from Figure 12 (middle row), we can see that, overall, the ECDI's frequency is decreasing with $E_y$ whereas the MTSI exhibit the opposite trend of increasing frequency as the $E_y$ increases. Also, the absolute values of the frequencies are different among various propellants such that the frequencies of the ECDI and the MTSI are the highest for Ar and the lowest for Xe in each $E_y$ case.

Finally, the azimuthal phase velocities are presented in Figure 12 (bottom row). These are calculated using $V_{ph} = \omega/k_z$, where $\omega$ is the frequency of the instability. It is noticed that the phase velocity of both the ECDI and the MTSI increases with the $E$-field intensity, with the strongest variation observed for Ar and the weakest variation for Xe. Moreover, in contrast to the MTSI, the phase velocity variation of the ECDI indicates a nearly linear relationship with $E_y$ within the studied range of $E$-field magnitudes.

### 3.3. Variation in the electrons' cross-field transport and the species' distribution function

In this section, we first examine, for the xenon propellant, the contribution of various terms in the electrons' azimuthal momentum equation to the cross-field transport. The approach followed here is similar to that used in Refs. [14][42] to isolate the various contributions in the axial-azimuthal simulations. The electron momentum equation along the azimuthal direction ($z$) can be written as

$$-qn_e v_{e,y} B_x = \partial_t(m_e n_e v_{e,z}) + \partial_x(m_e n_e v_{e,x} v_{e,z}) + \partial_x(\Pi_{e,xz}) - q\tilde{n}_e \tilde{E}_z \quad \text{(Eq. 3)}$$

where, $q$ is the unit charge, $n_e$ is the electron number density, $v_{e,x}$ and $v_{e,z}$ are the electron axial and azimuthal drift velocity, $B_x$ is the radial magnetic field intensity, and $\tilde{n}_e$ and $\tilde{E}_z$ are the fluctuating electron number density and azimuthal electric field. In this equation, the left-hand-side term is called the magnetic force ($F_B$) whereas, on the right-hand-side, the first term is the temporal inertia force ($F_t$), the second term is the convective inertia force ($F_I$), the third force term corresponds to the viscous effects ($F_\Pi$), and the last term is the electric force term ($F_E$). We discussed in Ref. [14] that the term $F_E$ represents the contribution of the azimuthal instabilities to the cross-field transport in our simulations.

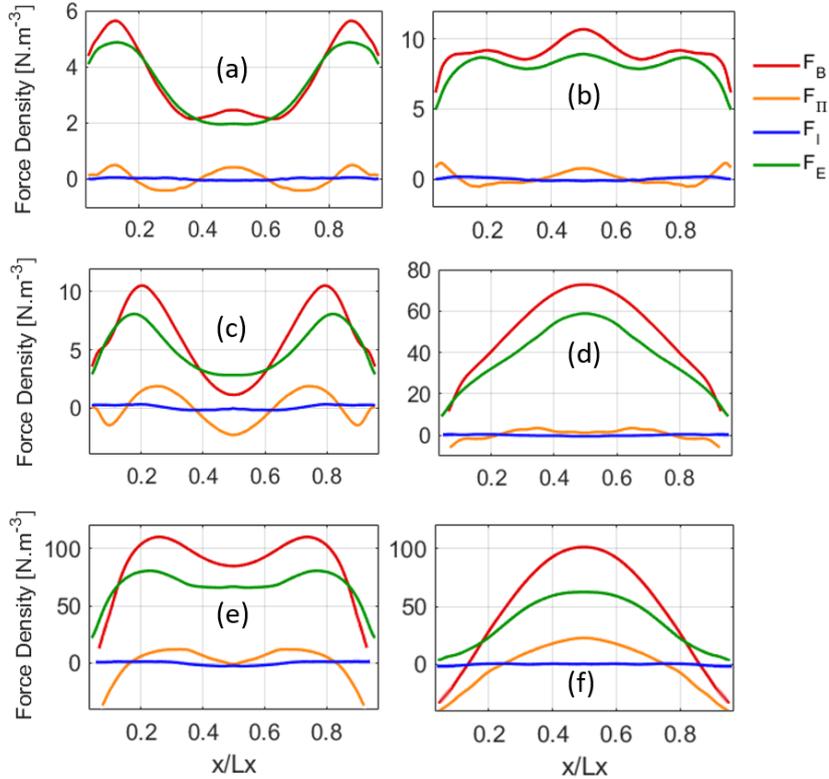

Figure 13: Radial distribution of the force terms in Eq. 3 for the xenon propellant and various values of the axial electric field: (a) $0.5E_0$, (b) $1E_0$, (c) $2E_0$, (d) $3E_0$, (e) $4E_0$, (f) $5E_0$. The momentum terms are averaged over 10 $\mu s$ of the simulations' time.

The radial distribution of the various force terms in Eq. 3 averaged over $10\ \mu s$ are presented in Figure 13 for various axial electric field intensities. Also, the radially averaged value of each transport term is plotted against the $E_y$ in Figure 14. The role of the $F_t$ term was seen to be negligible and, therefore, it is not shown in the plots. From these figures, it is evident that the $F_E$ term constitutes the major contribution in all simulation cases. Therefore, the $F_E$ profile determines the overall distribution of the cross-field transport across the domain. The radial distribution of the $F_E$ term in each case resembles the respective profile of the $J_{ey}$ in Figure 4 for xenon. The second important force term is $F_\Pi$, whose magnitude is almost one order of magnitude less than $F_E$. For lower $E$-field values ($0.5E_0$, $1E_0$, and $2E_0$), this term has an oscillatory profile around zero along radial direction, resulting in a mean value that is close to zero. For larger $E$-fields ($3E_0$, $4E_0$, and $5E_0$), the $F_p$ term remains positive over an extended portion of the domain around the center and become negative near the walls. Furthermore, our results show that the convective inertia term ($F_I$) has a negligible role in cross-field transport across all $E_y$ cases.

It is noteworthy from Figure 14 that, in some simulation cases, especially in the presence of larger $E$-fields, the sum of all terms on the right-hand side of Eq. 3 (RHS) becomes less than the $F_B$ term on the left-hand side (LHS). This discrepancy is a consequence of the application of the boundary condition on the particles' motion along the axial direction, which served to prevent the particles' energy from indefinitely increasing. In this regard, as explained in Section 2, particles crossing the artificial boundaries of the domain along the axial direction are reloaded on the simulation plane with new velocities sampled from their initial distribution functions. This resampling of the particles amounts to a momentum loss which manifests itself as a difference between the RHS and the LHS of the Eq. 3.

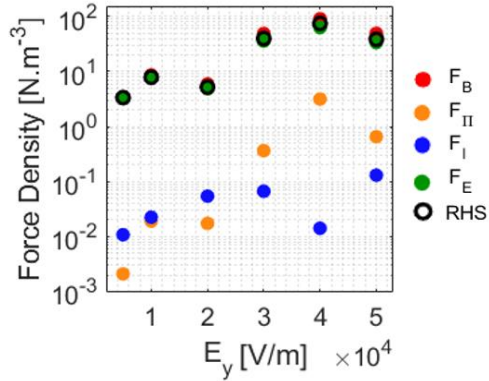

Figure 14: Variation vs $E_y$ of the radially averaged value of each force term in the electrons' azimuthal momentum equation (Eq.3) from the radial-azimuthal simulations with xenon propellant. The hollow black dot represents the sum of all terms on the right-hand-side of Eq.3.

We have presented the normalized velocity distribution functions of the electrons (Figure 15) and the ions (Figure 16) along the radial and the azimuthal velocity components at the end of the simulations, i.e., at t = 30 $\mu s$, for various propellants and $E_y$ values.

From Figure 15, we notice that, at the low $E$-field magnitudes ($0.5E_0$, $1E_0$, and $2E_0$), the tail of the radial EVDFs are mostly depleted, which is consistent with the arguments made in Section 3.1 regarding the existence of strong MTSI in this $E_y$-intensity range, its consequence on the electrons' heating along the radial direction, and the correlation with the significance of the electron particles flux to the wall. As the $E$-field value increases, the MTSI becomes weaker, and the ECDI dominates, a larger portion of the electrons in the tail of the distribution functions are consistently retained. In addition, in the $3E_0$ case, a slight depletion of particles at the mid radial-velocity range can be observed, which is more significant for Ar and Kr. The radial EVDFs in the $4E_0$ and $5E_0$ conditions are the broadest for Xe and the narrowest for Ar, observations that are consistent with the relative $T_{ex}$ in these simulations according to Figure 2.

Looking at the azimuthal EVDFs, it is seen that the distribution functions shift toward negative velocities as $E$-field value increases, showing expectedly larger electrons' azimuthal drift. Also, the distribution functions extend over a progressively broader velocities when the $E$-field increases, which is in line with the trend of the $T_{ez}$ observed in Figure 2(b) and (e).

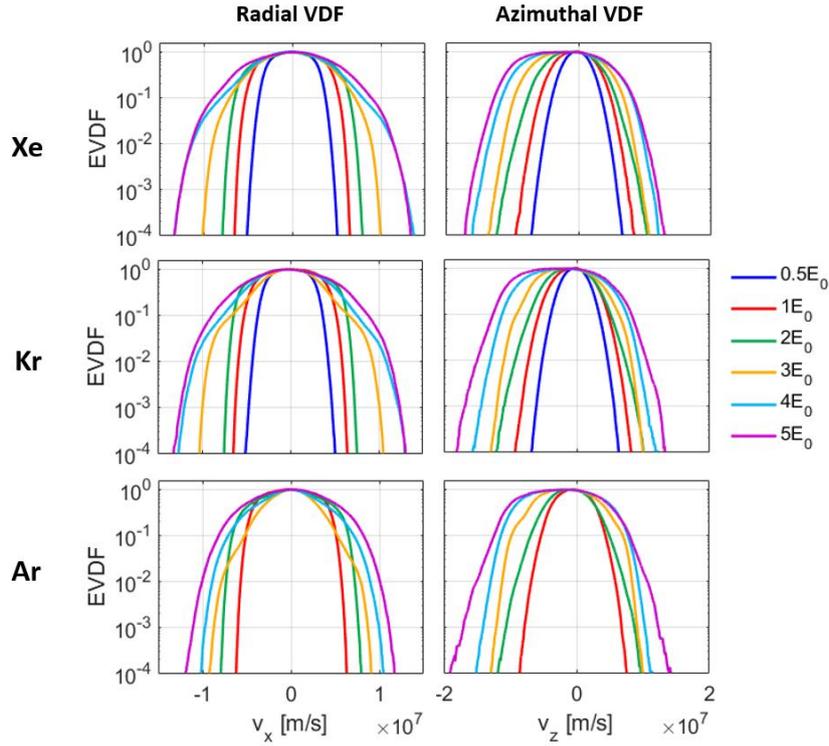

Figure 15: Normalized radial and azimuthal velocity distribution functions of the electrons for various axial electric field values and the three studied propellants from the quasi-2D simulations.

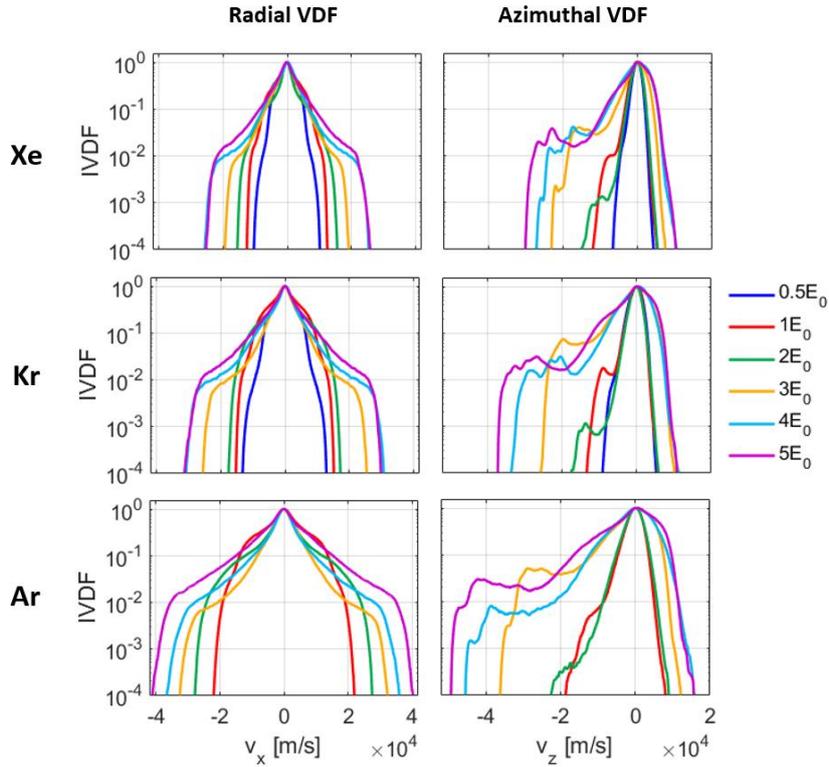

Figure 16: Normalized radial and azimuthal velocity distribution functions of the ions for various axial electric field values and the three studied propellants from the quasi-2D simulations.

Regarding the ions' velocity distribution functions (Figure 16), it is noticed that the radial IVDFs experience significant broadening for larger $E$-fields, especially in the $3E_0$, $4E_0$ and $5E_0$ conditions. The radial IVDFs are also depleted in the mid radial-velocity range.

Moreover, the azimuthal IVDFs indicate increasing broadening toward negative velocities with $E_y$, which is due to the interaction of the ion particles with the azimuthal waves. The IVDF broadening progressively increases from $0.5E_0$ case to $5E_0$ case as a result of the ions interacting with azimuthal waves of increasingly larger phase velocities according to Figure 12 (bottom row). It is also noticed that Ar has the broadest and Xe has the narrowest IVDFs along both the radial and the azimuthal directions in each $E_y$ case, which is again due to the higher phase velocities of the waves in case of Ar.

**Section 4: Conclusions**

We investigated in this paper the effect of the axial electric field intensity on E×B plasma discharges of various gases, xenon, krypton, and argon. The adopted simulation setup corresponded to a radial-azimuthal cross-section of a Hall thruster for which the excitation and presence of two main instability modes, the ECDI and the MTSI, had been reported in previous works [16][17][31][33]. Most notably from the present effort, we demonstrated that the magnitude of the axial electric field delimits two distinct regimes of the plasma behavior corresponding to the dominance of either the ECDI or the MTSI. It was shown across all studied propellants that, for relatively low values of the axial electric field ($E_y \leq 20\ kVm^{-1}$), the MTSI is the dominant instability mode. For relatively high values of the $E_y$, i.e., $E_y \geq 30\ kVm^{-1}$, the MTSI becomes mitigated, giving room to the dominance of the ECDI. We showed that the critical $E_y$ value at which the change in the plasma regime occurs is about $25\ kVm^{-1}$. The variation in the axial electric field value and the consequent change in the plasma regime was underlined to alter the plasma properties distribution and the characteristics of the underlying phenomena.

In terms of the time-averaged plasma properties, increasing $E_y$ was seen to result in lower peaks of the ion number density for all propellants. The radially averaged total electron temperature as well as the azimuthal and radial temperature components were observed to increase with the $E_y$. However, it was demonstrated that the decrease in the peak value of the ion number density and the increase in the mean radial electron temperature are not monotonic vs $E_y$. In fact, aligned with the change in the plasma regime at $E_y \approx 25\ kVm^{-1}$, the decreasing trend of peak $n_i$ and the increasing trend of mean $T_{ex}$ undergo a deviation from monotonic behavior, with the former presenting a jump in values and the latter exhibiting kind of a plateau.

The electron temperature anisotropy, represented as the ratio of the radial-to-azimuthal temperatures ($T_{ex}/T_{ez}$), was noticed to be a perfect measure to distinct the two plasma regimes for all propellants. Indeed, due to the notable radial heating effect of the MTSI, in the regime where this instability is dominant, $T_{ex}/T_{ez}$ becomes close to or larger than 1 depending on the specific propellant. On the contrary, where ECDI is dominant and MTSI mitigated, $T_{ex}/T_{ez}$ becomes notably smaller than 1 since the ECDI mostly heats up the electron population along the azimuthal direction. As the dominant instability mode changes, the radial profile of the axial electron current density varies accordingly such that, in the cases with dominant MTSI, the profiles peak near the walls whereas, in the mitigated MTSI cases, the $J_{ey}$ profile peaks around the center of the domain. We observed that, in case the MTSI and the ECDI coexist with rather similar magnitudes, the $J_{ey}$ radial profile tends to become uniform along the radial extent of the domain.

We verified our observations regarding the variation in the relative strength of the developed instabilities in the discharge with $E_y$ magnitude within distinct plasma regimes through extensive spectral analyses. We employed spatial and temporal FFTs in addition to the Optimized Dynamic Mode Decomposition method, which we recently demonstrated as a powerful tool for simultaneous spatial-temporal characterization of the instabilities and fluctuations in plasma systems [38]. We compared the change in the simulated azimuthal wavenumber of the dominant ECDI and the MTSI modes vs $E_y$ against the available theories from the literature, observing great agreement between the simulations' predictions and the theoretical values. We also assessed, for various propellants, the variation in the frequency and azimuthal phase velocity of the dominant ECDI and MTSI modes as the $E_y$ value changes. It was highlighted that, although the phase velocity of the ECDI and the MTSI both increases with $E_y$, the frequency variation trends are opposite between the two instabilities. For the ECDI, the frequency of the first harmonic was shown to overall decrease with increasing $E_y$, whereas, for the MTSI, the frequency exhibits an increasing trend vs axial electric field intensity.

In terms of the contribution to electron transport of various force terms in the electrons' azimuthal momentum equation, we presented analyses for xenon as an example, observing that, across all values of $E_y$, the dominant role is played by the electric force term due to the influence of the correlated number density and electric field fluctuations. The viscous force term due to the off-diagonal terms in the electrons' pressure tensor were seen to also play a non-negligible role for high $E_y$ values, particularly $4E_0$ and $5E_0$. Interestingly, the radial distributions of the viscous force term were noticed to vary with the change in $E_y$ and in a manner consistent with the identified distinct plasma regimes.

The electrons' and ions' velocity distribution functions along the radial and azimuthal velocity components were analyzed for our simulation cases with different $E_y$ values and various propellants. Consistent with the change in plasma regime and the dominance of different instability modes, the electrons' radial VDF was particularly seen to be modified. In cases with dominant MTSI, the VDFs' tail was noticed to be depleted whereas, when the MTSI is mitigated, more tail electrons were retained. The ions' radial and azimuthal velocity distribution functions also exhibited notable distortion and broadening, which were noticed to become more pronounced for high values of $E_y$ and for lighter propellants.

Finally, to place the findings of this article in an applied context, it is important to note that, in a realistic setting, for example during the operation of a Hall thruster, the plasma discharge features a back-and-forth motion, alongside which the peak magnitude of the axial electric field periodically changes [1][18]. In this regard, our results in this paper, which are subject to further verification in a 3D simulation setup with self-consistent resolution of the axial direction as well as the ionization process, indicate that this periodic variation in the peak electric field intensity may notably affect the underlying instability mechanisms, influencing the significance of plasma species' heating and transport.

**Acknowledgments**:

The present research is carried out within the framework of the project "Advanced Space Propulsion for Innovative Realization of space Exploration (ASPIRE)". ASPIRE has received funding from the European Union's Horizon 2020 Research and Innovation Programme under the Grant Agreement No. 101004366. The views expressed herein can in no way be taken as to reflect an official opinion of the Commission of the European Union.

The authors gratefully acknowledge the computational resources and support provided by the Imperial College Research Computing Service (http://doi.org/10.14469/hpc/2232).

**Data Availability Statement**:

The simulation data that support the findings of this study are available from the corresponding author upon reasonable request.

**References**:

[1] Boeuf JP, "Tutorial: Physics and modeling of Hall thrusters", *Journal of Applied Physics* **121** 011101 (2017)

[2] Boeuf JP, Garrigues L, "E × B electron drift instability in Hall thrusters: Particle-in-cell simulations vs. theory", *Phys. Plasmas* **25**, 061204 (2018)

[3] Taccogna F and Garrigues L, "Latest progress in Hall thrusters plasma modelling", *Reviews of Modern Plasma Physics*, Springer Singapore, 3 (**1**) (2019)

[4] Kaganovich ID, Smolyakov A, Raitses Y, et al., "Physics of E × B discharges relevant to plasma propulsion and similar technologies", *Phys. Plasmas* **27**, 120601 (2020)

[5] Boeuf JP, Smolyakov A, "Physics and instabilities of low-temperature E × B plasmas for spacecraft propulsion and other applications", *Physics of Plasmas* **30**, 050901 (2023)

[6] Reza M, Faraji F, Knoll A, "Concept of the generalized reduced-order particle-in-cell scheme and verification in an axial-azimuthal Hall thruster configuration", *J. Phys. D: Appl. Phys.* **56** 175201 (2023)

[7] Juhasz Z, Ďurian J, Derzsi A, Matejčík S, Donkó Z, Hartmann P, "Efficient GPU implementation of the Particle-in-Cell/Monte-Carlo collisions method for 1D simulation of low-pressure capacitively coupled plasmas", *Computer Physics Communications*, **263**, 107913 (2021)

[8] Eremin D, "An energy- and charge-conserving electrostatic implicit particle-in-cell algorithm for simulations of collisional bounded plasmas", *Journal of Computational Physics*, **452**, 110934 (2022)

[9] Taccogna F, Minelli P, "Three-dimensional particle-in-cell model of Hall thruster: The discharge channel", *Physics of Plasmas* **25**, 061208 (2018)

[10] Garrigues L, Tezenas du Montcel B, Fubiani G et al., "Application of sparse grid combination techniques to low temperature plasmas particle-in-cell simulations. I. Capacitively coupled radio frequency discharges", *J. Appl. Phys.* **129**, 153303 (2021)

[11] Birdsall CK, Langdon AB, "Plasma Physics via Computer Simulation", CRC Press, 1991

[12] Charoy T, "Numerical study of electron transport in Hall thrusters", Plasma Physics [physics.plasm-ph], Institut Polytechnique de Paris. English. ⟨NNT: 2020IPPAX046⟩ ⟨tel-02982367⟩ (2020)

[13] Reza M, Faraji F, Knoll A, "Generalized reduced-order particle-in-cell scheme for Hall thruster modeling: concept and in-depth verification in the axial-azimuthal configuration", ArXiv pre-print, DOI: arXiv:2208.13106 (2022)

[14] Faraji F, Reza M, Knoll A, "Enhancing one-dimensional particle-in-cell simulations to self-consistently resolve instability-induced electron transport in Hall thrusters". *J. Appl. Phys.* **131**, 193302 (2022)

[15] Reza M, Faraji F, Knoll A, "Resolving multi-dimensional plasma phenomena in Hall thrusters using the reduced-order particle-in-cell scheme", *J Electr Propuls* **1**, 19 (2022)

[16] Faraji F, Reza M, Knoll A, "Verification of the generalized reduced-order particle-in-cell scheme in a radial-azimuthal E×B plasma configuration", *AIP Advances* **13**, 025315 (2023)

[17] Reza M, Faraji F, Knoll A, "Parametric investigation of azimuthal instabilities and electron transport in a radial-azimuthal E×B plasma configuration", *Journal of Applied Physics* **133**, 123301 (2023)

[18] Reza M, Faraji F, Knoll A, Piragino A, Andreussi T, Misuri T, "Reduced-order particle-in-cell simulations of a high-power magnetically shielded Hall thruster", *Plasma Sources Sci. Technol.* **32**, 065016 (2023)

[19] Fossum EC, King LB, "Design and Construction of an Electron Trap for Studying Cross-Field Mobility in Hall Thrusters", AIAA 2007-5207, 43rd AIAA/ASME/SAE/ASEE Joint Propulsion Conference & Exhibit, Cincinnati, Ohio (2007)

[20] Ekholm JM, Hargus, Jr. WA, "E×B Measurements of a 200 W Xenon Hall Thruster", AIAA-2005-4405, 41st AIAA/ASME/SAE/ASEE Joint Propulsion Conference & Exhibit, Tucson, Arizona (2005)

[21] Raitses Y, Kaganovich D, Khrabrov A, Sydorenko D, Fisch NJ, Smolyakov A, "Effect of Secondary Electron Emission on Electron Cross-Field Current in E×B Discharges", in *IEEE Transactions on Plasma Science*, vol. **39**, no. 4, pp. 995-1006 (2011)

[22] Brennan MJ, Garvie AM, "An experimental investigation of electron transport in E×B discharges", *Australian Journal of Physics*, vol. **43**, p.765 (1990)

[23] Goebel DM and Katz I, "Fundamentals of Electric Propulsion: Ion and Hall Thrusters", Wiley, 2008. DOI: 10.1002/9780470436448

[24] Janhunen S, Smolyakov A, Sydorenko D, Jimenez M, Kaganovich I, Raitses Y, "Evolution of the electron cyclotron drift instability in two dimensions," *Phys. Plasmas* **25**, 082308 (2018)

[25] Croes V, Lafleur T, Bonaventura Z, Bourdon A, Chabert P, "2D particle-in-cell simulations of the electron drift instability and associated anomalous electron transport in Hall-effect thrusters", *Plasma Sources Sci. Technol.* **26** 034001 (2017)

[26] A. Ducrocq, J. C. Adam, A. Héron, and G. Laval, "High-frequency electron drift instability in the cross-field configuration of Hall thrusters", *Phys. Plasmas*, **13** 102111 (2006)

[27] Lazurenko A, Vial V, Bouchoule A., Prioul M, et al, "Characterization of microinstabilities in Hall thruster plasma: experimental and PIC code simulation results", IEPC-2003-0218, *In Proceedings of International Electric Propulsion Conference*, Toulouse, France (2003)

[28] Cavalier J, Lemoine N, Bonhomme G, Tsikata S, Honore C, Gresillon D, "Hall thruster plasma fluctuations identified as the E×B electron drift instability: Modeling and fitting on experimental data", *Phys. Plasmas* **20**, 082107 (2013)

[29] Tsikata S, Honore C, Lemoine N, Gresillon DM, "Three-dimensional structure of electron density fluctuations in the Hall thruster plasma: the E×B mode", *Phys. Plasmas* **17**, 112110 (2010)

[30] Tsikata S, Honore C, Gresillon DM, "Collective Thomson scattering for studying plasma instabilities in electric thrusters", *J. Instrum.* **8**, C10012 (2013)

[31] Petronio F, Tavant A, Charoy T, Alvarez-Laguna A, Bourdon A, Chabert P, "Conditions of appearance and dynamics of the Modified Two-Stream Instability in E×B discharges", *Phys. Plasmas* **28**, 043504 (2021)

[32] Janhunen S, Smolyakov A, Chapurin O, Sydorenko D, Kaganovich I, Raitses Y, "Nonlinear structures and anomalous transport in partially magnetized E×B plasmas," *Phys. Plasmas* **25**, 011608 (2018)

[33] Villafana W, Petronio F, Denig AC, Jimenez MJ, et al., "2D radial-azimuthal particle-in-cell benchmark for E×B discharges", *Plasma Sources Sci. Technol.* **30** 075002 (2021)

[34] Koshkarov O, Smolyakov A, Raitses Y, Kaganovich I, "Self-organization, structures, and anomalous transport in turbulent partially magnetized plasmas with crossed electric and magnetic fields," *Phys. Rev. Lett.* **122**, 185001 (2019)

[35] Petronio F, "Plasma instabilities in Hall Thrusters: a theoretical and numerical study", PhD dissertation, Paris Polytechnic Institute, NNT: 2023IPPAX030 (2023)

[36] Croes V, Tavant A, Lucken R, Martorelli R, Lafleur T, Bourdon A, Chabert P, "The effect of alternative propellants on the electron drift instability in Hall-effect thrusters: Insight from 2D particle-in-cell simulations", Physics of Plasmas 25, 063522 (2018)

[37] Lafleur T, Baalrud SD, Chabert P., "Theory for the anomalous electron transport in Hall effect thrusters. I. Insights from particle-in-cell simulations," *Phys. Plasmas* **23**, 053502 (2016)

[38] Faraji F, Reza M, Knoll A, Kutz JN, "Dynamic Mode Decomposition for data-driven analysis and reduced-order modelling of E×B plasmas: I. Extraction of spatiotemporally coherent patterns", ArXiv Preprint, arXiv:2308.13726 (2023)

[39] Adam JC, Heron A, Laval G, "Study of stationary plasma thrusters using two-dimensional fully kinetic simulations", *Phys. Plasmas* **11**, 295 (2004)

[40] Taccogna F, Minelli P, Asadi Z, Bogopolsky G, "Numerical studies of the E × B electron drift instability in Hall thrusters," *Plasma Sources Sci. Technol.* **28**, 064002 (2019)

[41] Askham T, Kutz JN, "Variable projection methods for an optimized dynamic mode decomposition", *SIAM Journal on Applied Dynamical Systems* **17**:1, 380-416 (2018)

[42] Lafleur T and Chabert P, "The role of instability-enhanced friction on 'anomalous' electron and ion transport in Hall-effect thrusters", *Plasma Sources Science and Technology* **27** 015003 (2018)